\documentclass[preprint2]{aastex}
\begin{document}
\title{Type II Cepheids as Extragalactic Distance Candles}
\author{Daniel J. Majaess, David G. Turner, David J. Lane}
\affil{Saint Mary's University, Halifax, Nova Scotia, Canada}
\affil{The Abbey Ridge Observatory, Stillwater Lake, Nova Scotia, Canada}
\email{dmajaess@ap.smu.ca}

\begin{abstract}
Extragalactic Type II Cepheids are tentatively identified in photometric surveys of IC 1613, M33, M101, M106, M31, NGC 4603, and the SMC.  Preliminary results suggest that Type II Cepheids may play an important role as standard candles, in constraining the effects of metallicity on Cepheid parameters, and in mapping extinction.  
\end{abstract}


\section{Introduction}
Type II Cepheids were often cited as potential distance indicators, yet the possibility of a non-unique period-luminosity relation and their comparative faintness relative to classical Cepheids often precluded their use.  Recent advances in the field are challenging such perceptions, however.  The OGLE survey of the LMC and Galactic bulge secured a statistically valid sample of Type II Cepheids that facilitated the establishment of viable period-magnitude relations \citep{ud99,ku03,so08}.  The distances inferred to the Galactic center and innumerable globular clusters from Type II Cepheids agree with estimates found in the literature \citep{pr03,mat06,gr08,fe08,mat09,ma09}.  Furthermore, Type II Cepheids were observed beyond the Local Group alongside classical Cepheids in the galaxy M106 \citep{ma06}.  That finding, in tandem with the possible discovery of Type II Cepheids in NGC 3198 \& NGC 5128 \citep{ke99,fe07,ma09}, motivated the present study which expands the search for extragalactic Type II Cepheids.

Type II Cepheids may be identified in galaxies that do not contain young massive stars like classical Cepheids \citep{tu96b}, since the variables originate from an older low mass population \citep{wa02}. It follows that statistical uncertainties in recent estimates of $H_0$ might be reduced through an increase in the number of galaxies used for calibration \citep{fm96,fr01}. Type II Cepheids also offer an empirical resolution to the debate surrounding the effect of metallicity on the  distances and colours of classical Cepheids.  The distance and reddening determined for a particular galaxy from classical Cepheids, Type II Cepheids, and RR Lyrae variables should be comparable, with any identifiable differences possibly linked to metallicity effects \citep{ma09}.  Similarly, parameters ($d$, $E_{B-V}$) established for globular clusters from RR Lyrae and Type II Cepheid variables may be compared to estimates that are insensitive to metallicity.   

\begin{figure}[!t]
\includegraphics[width=7.5cm]{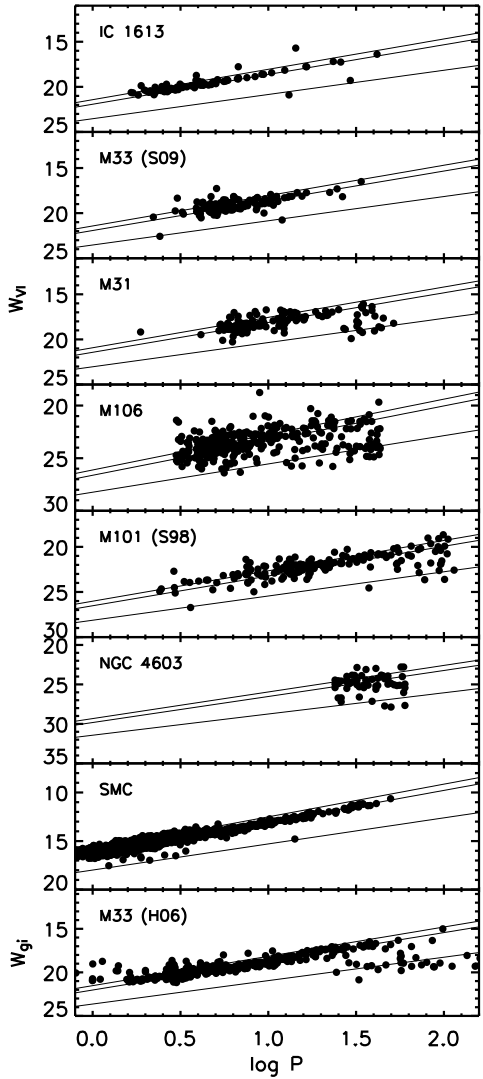}
\caption{\small{Wesenheit period-magnitude diagrams for the galaxies studied.  Solid lines indicate the Wesenheit functions for classical Cepheids pulsating in the overtone and fundamental modes, and Type II Cepheids.}}
\label{fig1}
\end{figure}

This study employs reddening-free Cepheid relations to highlight the potential membership of Type II Cepheids in surveys of IC 1613 \citep{ud01}, M106 \citep{ma06}, M33 \citep{ma01,ha06,be08,sc09}, M101 \citep{ke96,st98}, M31 \citep{bo03}, NGC 4603 \citep{ze97,ne99}, and the SMC \citep{ud99}.  The effects of metallicity on the determination of a Cepheid's colour and distance are also discussed. 

\section{Extragalactic Type II Cepheids}
\label{dist}
Wesenheit reddening-free parameterizations for classical Cepheids pulsating in the first overtone (OT) and fundamental mode (FU), and for Type II Cepheids (TII) can be deduced from OGLE observations of the LMC \citep{ud99,so08}:
\begin{eqnarray}
\label{eqn0}
W_{VI}=V-\beta(V-I) \nonumber \\
W_{VI}(FU)= -(3.29\pm0.02)(\log{P})+(15.82\pm0.01)  \nonumber \\
W_{VI}(OT)= -(3.37\pm0.02)(\log{P})+(15.32\pm0.01) \nonumber \\
W_{VI}(TII^*)\simeq -2.7\times\log{P}+17.4 \nonumber \\
\end{eqnarray}
BL Her, W Vir, and RV Tau variables do not follow the same simple Wesenheit function \citep[see][]{so08}.  Consequently, the aforementioned equations were only used to identify Type II and classical Cepheids by their locations in reddening-free period-magnitude diagrams \citep{vb68,ma82,op83}.  The distances were then computed from the reddening-free Cepheid distance relationship of \citet{ma08,ma09}.   A correction term was derived by \citet{ma09} to permit the determination of distances to the RV Tau subclass of Type II Cepheids, in addition to stars occupying the BL Her and W Vir regimes.  The reddening-free Cepheid distance relationships were obtained via least squares techniques applied to specific samples of calibrators. For classical Cepheids the sample consisted of Galactic cluster Cepheids \citep[e.g.,][]{tu02} and Cepheids with new HST parallaxes \citep{be07}. Defining the relation strictly as a Galactic calibration is somewhat ambiguous given that Milky Way Cepheids appear to follow a galactocentric metallicity gradient \citep{an02,an02b}. The \citet{ma08} relation is tied to Galactic classical Cepheids that exhibit near solar abundances \citep{an02,an02b}. For Type II Cepheids the calibrators were LMC variables observed by OGLE \citep{ud99,so08}.

The colour coefficient for the Wesenheit relations used here, $\beta=-2.55$, is that employed by \citet{fo07}.  The slopes of the $W_{VI}$ relations do vary between galaxies, but are not large enough to affect the present objectives and shall be elaborated upon in a separate study.  The Wesenheit functions were shifted in tandem by the same zero-point to correct for a difference in modulus between the calibrating galaxy (the LMC) and the target galaxy under inspection.  It is noted that the formal uncertainties cited for the Wesenheit functions are optimistic (Eqn.~\ref{eqn0}).  

Classical and Type II Cepheid candidates in the survey of M33 by \citet{ha06} and \citet{be08} were identified by constructing a Wesenheit function for $g'$ and $i'$ magnitudes. A sizeable Type II Cepheid population exists in that survey, of which a small subsample are plotted in Fig.~\ref{fig1}.   Further results by their research group are eagerly anticipated.

\begin{figure}[!t]
\includegraphics[width=6.5cm]{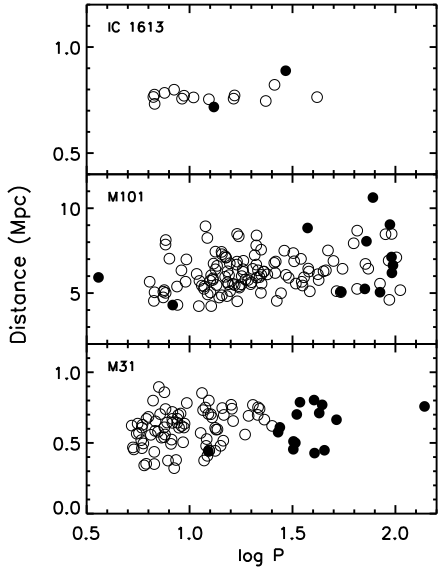}
\caption{\small{Cepheid period-distance diagrams for the galaxies IC 1613, M101, and M31.  Filled circles identify stars analyzed using the Type II Cepheid distance relation. Open circles identify stars analyzed with the classical Cepheid distance parameterization.}}
\label{fig2}
\end{figure}

Type II Cepheid candidates were tentatively identified in IC 1613 \citep{ud01}, M106 \citep{ma06}, M33 \citep{ma01,ha06,be08,sc09}, M101 \citep{ke96,st98}, M31 \citep{bo03}, NGC 4603 \citep{ze97,ne99}, and the SMC \citep{ud99}.  Data for a subsample of Type II Cepheid candidates are presented in Table~\ref{arophot} as an appendix.  Once classifications were established from a Wesenheit analysis (Fig.~\ref{fig1}), period-distance diagrams were constructed using reddening-free $VI$ classical and Type II Cepheid distance relations \citep{ma08,ma09}.  A subsample is provided as Fig.~\ref{fig2}.  Parameters for the galaxies are summarized in Table \ref{dgalaxies}, where (i) and (o) denote the inner and outer regions of the galaxies. Inadequate sampling is a concern and may result in systemic offsets whenever small statistics are present (IC1613, SMC, M33).   The SMC sample may be biased by anomalous Cepheids or the limiting magnitude of the survey. The distance was therefore weighted towards stars near the center of the Type II Cepheid relation.

\begin{deluxetable}{llccc}
\tabletypesize{\scriptsize}
\tablecaption{Distance moduli for the sample galaxies.\label{dgalaxies}} 
\tablewidth{0pt}
\tablehead{\colhead{Galaxy} &\colhead{$(m-M)_0$ (TI)} &\colhead{$(m-M)_0$ (TII)} &\colhead{No. TII} &\colhead{Photometry}}
\startdata
IC 1613 &$24.35\pm0.09$ &$24.52\pm0.16$ &2 &1 \\
SMC &$18.93\pm0.10$ &$18.85\pm0.11$\tablenotemark{a} &9 &2 \\
M33 &$24.43\pm0.14$ (i) &$24.54$ &1 &3 \\
&$24.67\pm0.07$ (o) &$\cdots$ &$\cdots$ &3 \\
&$24.40\pm0.17$ (i) &$24.5\pm0.3$ & 5 & 4 \\
M31 (Y-field) &$23.93\pm0.24$ &$23.93\pm0.24$ &17 &5 \\
M106 &$29.09\pm0.14$\tablenotemark{a} (i) &$29.20\pm0.18$ &15 &6 \\
&$29.20\pm0.12$\tablenotemark{a} (i) &$\cdots$ &$\cdots$ &6 \\
&$29.34\pm0.09$ (o) &$29.43\pm0.14$ &6 &6 \\
&$29.46\pm0.16$ (i) &$\cdots$ &$\cdots$ &7,8 \\
M101 &$28.89\pm0.17$ (i) &$28.9\pm0.4$ &6 &9 \\
&$29.29\pm0.20$ (o) &$\cdots$ &$\cdots$ &10 \\
NGC 4603 &$32.3\pm0.4$ &$31.6\pm0.3$ &5 &7,11 \\ 
\enddata
\tablecomments{References: (1) \citet{ud01}, (2) \citet{ud99}, (3) \citet{sc09}, (4) \citet{ma01}, (5) \citet{bo03}, (6) \citet{ma06}, (7) \citet{new01}, (8)\citet{mao99}, (9) \citet{st98}, (10) \citet{ke99}, (11) \citet{ze97}.}
\tablenotetext{a}{see text}
\end{deluxetable}

\citet{so08} discovered a set of 16 rather interesting and peculiar LMC Type II Cepheids of the W Vir subclass. The stars exhibit periods of pulsations ranging from $\simeq4-10$ days, and were interpreted as more luminous binary systems \citep{so08}.   Indeed, the enigmatic IX Cas may be the Galactic analog \citep{hw89,tu09c}. Such stars occupy a small fraction ($\sim 10 \%$) of the overall Type II Cepheid population in the LMC, mitigating the impact on conclusions derived here owing to misclassified W Vir pulsators. However, depending on the limiting magnitude of the survey, caution is warranted since such stars may be preferentially sampled owing to their increased luminosities relative to regular W Vir pulsators.

Admittedly, stars highlighted in Table~\ref{arophot} may be semi-regulars or variables of differing classes that overlap the Wesenheit relation describing Type II Cepheids \citep{so07,so08,so09,pe09}.  An identification scheme based solely on a variable's position within the Wesenheit diagram is inadequate (Fig.~\ref{fig1}).  The initial sample presented in the Wesenheit and period-distance diagrams (Fig.~\ref{fig1},~\ref{fig2}) were subsequently purged of variables exhibiting apparent colours significantly redder than the Cepheids.  Semi-regulars, for example, are typically redder and may be magnitudes fainter than Cepheids \citep{ud99,so07,so08,so09}.  Classical and Type II Cepheids exhibit similar colours at a given period, but diverge in particular towards longer periods where RV Tau stars appear bluer than classical Cepheids.  Applying the period-color diagnostic reduced the number of Type II Cepheid candidates by $\simeq50$\%.  No account for differential reddening was made.   Additional work is needed here, as a rigorous analysis based on complete data should include a multifaceted approach to highlight contaminators using: Fourier parameters for the light-curves, color-magnitude, period-amplitude, period-color, and Wesenheit diagrams.

\section{Cepheid Metallicity Effect}
\subsubsection{Effects on color}
\label{smetallicity}
Precise metallicity independent distance estimates for the sample of galaxies studied here are rare, for example the maser distance to M106 \citep{he05}. That complicates efforts to constrain the effect of metallicity on Cepheid distances. Conversely, reddenings are readily available and established by a set of autonomous methods. The intrinsic colour of a Cepheid, and hence its temperature, are considered sensitive to metallicity \citep[see references in the review by][]{fe99}. Consequently, a classical Cepheid $VI$ colour-excess relation (\S \ref{sextinction}, Eqn.~\ref{eqn1}) calibrated with Galactic variables in the solar neighborhood should yield a spurious estimate of reddening for classical Cepheids in the metal-poor galaxy IC 1613 \citep{ud01}. The two samples exhibit a sizeable metallicity difference, namely $\Delta[Fe/H]\simeq1$. However, the reddening of classical Cepheids in IC 1613 established from Eqn.~\ref{eqn1} agrees with that obtained by metallicity independent means (see Table~\ref{reds}).   It follows that to within the uncertainties a classical Cepheid's intrinsic $(V-I)$ colour must be relatively insensitive to metallicity.  The colour-excess relation (Eqn.~\ref{eqn1}) also appears to provide a reasonable estimate of reddening for Type II Cepheids (RV Tau subclass excluded), and an additional test for metallicity effects. That is demonstrated by the good agreement of reddenings for Type II Cepheids in the globular clusters NGC 6441 \citep{pr03}, M54 \citep{pr03}, M15 \citep{co08}, and the LMC with metallicity independent determinations (Table~\ref{reds}).  No metallicity corrections were applied despite a sizeable range in abundance spanning $[Fe/H]\sim-0.3$ to $-2.3$ (Table~\ref{reds}).   A unique period-reddening relation for Type II Cepheids shall be developed in a follow-up study since Eqn.~\ref{eqn1} inadequately characterizes the entire period regime.   Nevertheless, the present results are confirmed by an ongoing parallel study that is establishing reddenings for a subsample of galaxies and globular clusters in Table~\ref{reds} from RR Lyrae variables.  The statistics are larger and shall bolster confidence in the results.  Lastly, the 2MASS reddenings highlighted in Table~\ref{reds} should be viewed as a first-order estimates. 

\begin{figure}[!t]
\includegraphics[width=6.5cm]{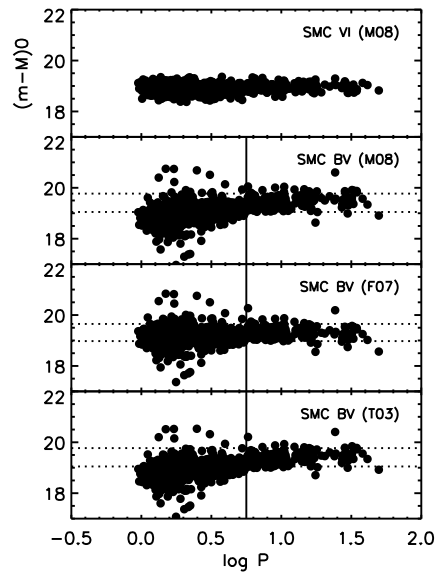}
\caption{\small{Cepheid period-distance diagrams for the SMC.  The slopes of classical Cepheid distance relations based on $BV$ photometry are sensitive to the effects of metallicity.  Conversely, the slopes of $VI$-based relations are relatively unaffected by comparison (top).  The Galactic classical Cepheid relations ($Z_{\odot}$) of \citet{ma08} (M08), \citet{fo07} (F07), and \citet{ta03} (T03) were used.}}
\label{fig3}
\end{figure}

Perhaps not surprisingly, the slope of the $VI$ reddening-free classical Cepheid distance relation also appears relatively insensitive to metallicity \citep{ud01,pi04,be07,vl07,fo07,ma08}. By contrast, the slope of a $BV$ relation is sensitive to metallicity.  Readers are referred to discussions in \citet{cc85}, \citet{ch93}, and \citet{ta03}.  The aforementioned trends are confirmed when computing the distance to SMC Cepheids from relationships calibrated with Galactic Cepheids in the solar neighborhood \citep[Fig.~\ref{fig3}, see also][]{ma08}. Classical Cepheids in the SMC are metal poor in comparison with Cepheids in the LMC and solar neighbourhood \citep{lu98,an02,an02b,mo06}.  $BV$-based Galactic classical Cepheid distance relations ineptly characterize SMC Cepheids (Fig.~\ref{fig3}), and a break in slope is apparent.  The ratio of total to selective extinction ($R$) can be adjusted to mitigate the bias noted in Fig.~\ref{fig3}, but the value required to linearize the $BV$ relation across all periods is unrealistic.  A disadvantage of the \citet{ma08,ma09} Cepheid parameterizations is the ratio of total to selective extinction is fixed. It is also noted, in hindsight, that the colour coefficient for the \citet{ma08} $BV$ relation may be rather large, although it is somewhat consistent with the value adopted by \citet{mf91}.  Further work is needed to consider the implications of anomalous values of $R$ \citep[e.g.,][]{ma01b,ud03}, and to examine why the \citet{ta03} and \citet{ma08} $BV$ relations do not match that of \citet{fo07} (Fig.~\ref{fig3}).  The latter finding is of particular concern, and a rigorous comparison of the Galactic calibrations shall ensue.

\subsubsection{Effects on distance}
\label{sdistance}
\begin{figure}[!t]
\includegraphics[width=6.5cm]{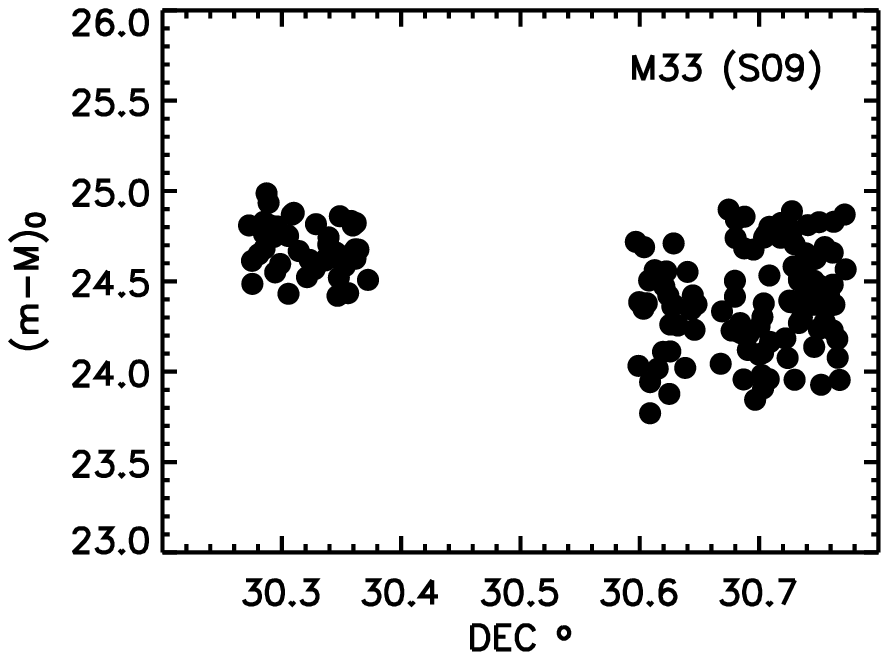}
\includegraphics[width=6.5cm]{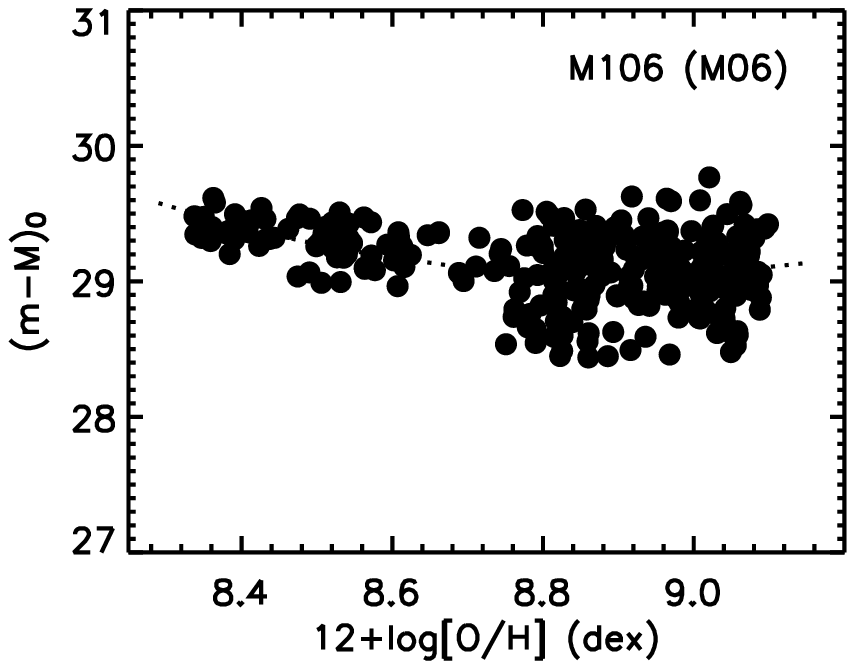}
\caption{\small{Cepheid declination-distance and abundance-distance diagrams for the galaxies M33 \citep[top, photometry from][]{sc09} \& M106 \citep[bottom, photometry from][]{ma06}.  Bottom, the data can be represented by a polynomial fit.}}
\label{fig4}
\end{figure}

The data in Table \ref{dgalaxies} indicate that distance moduli for the inner regions of spiral galaxies in the sample derived from Cepheids are consistently smaller than the outer regions. The canonical explanation attributes the difference to a metallicity gradient (Fig.~\ref{fig4}, bottom). However, there are concerns regarding that assertion. For example, the distances computed to the two data sets that sample the inner regions of M106 \citep{new01,ma06}, where the abundances are similar, differ by $(m-M)_0\simeq0.32$ (see Table \ref{dgalaxies}). The discrepancy may reflect the difficulty of achieving a common photometric zero-point and the need to reassess the error budget assigned to extragalactic Cepheid distance determinations.  The distance to classical Cepheids in the outer regions of M33 as established from the photometry of \citet{ma01} and \citet{sc09} is in excellent agreement (Table~\ref{dgalaxies}).  Likewise, the results highlighted in Table~\ref{dgalaxies} for the inner and outer regions of M101 are consistent with that cited in \citet{st98}.  The implied difference in distance between the two regions of M33 and M101 are: $\Delta (m-M)_{0,M33} \simeq 0.24$ and $\Delta (m-M)_{0,M101} \simeq 0.40$.   The observations by \citet{ma06} for M106 imply a difference of $\Delta (m-M)_{0,M106} \simeq 0.25$ (Fig.~\ref{fig4}, $P\ge7^d$), with the caveat that the results for M106 by \citet{mao99} and \citet{new01} are omitted.  Applying a period-cut ($P\ge12^d$) to the inner region's sample as indicated by \citet{ma06} reduces the offset to $\Delta (m-M)_{0,M106} \simeq 0.14$ \citep[see also Fig.17,][]{ma06}, however, an anomalous $VI$ Wesenheit slope remains and is a concern.  Nevertheless, evaluating $\delta (m-M)_0 / \delta [O/H]$ by straight division rather than applying a linear fit to the data, an analysis of M106 indicates that the functional dependence of metallicity on distance appears more aptly characterized as non-linear (e.g., a polynomial, Fig.~\ref{fig4}), yields $\gamma \sim 0.5$ mag dex$^{-1}$.  The required metallicity correction appears too large to account for the offset between the inner and outer regions of the galaxies examined.   

Galactic classical Cepheids provide a metallicity-uncorrected distance to the LMC, SMC, and IC 1613 of $(m-M)_0\simeq18.45,18.93,24.35$ \citep[e.g.,][and Table~\ref{dgalaxies}]{ma08}.  Applying a more modest metallicity correction of $\gamma \simeq 0.3$ mag dex$^{-1}$ reduces the distance to the LMC, SMC, and IC 1613 to $(m-M)_0 \simeq 18.35, 18.70, 24.05$.   The latter value for IC 1613 is especially disconcerting.  However, yet again IC 1613 provides a unique opportunity to verify the proposed metallicity effect owing to the sizeable abundance difference between classical Cepheids in the Milky Way and that galaxy ($\Delta[Fe/H]\simeq1$, see \S~\ref{smetallicity}).  Consider the following tests comparing distances to RR Lyrae variables, Type II Cepheids, and classical Cepheids at a common zero-point (e.g., IC 1613 \& the SMC).  \citet{be02} obtained a parallax for RR Lyrae using the Hubble Space Telescope, which when averaged with the Hipparcos estimate yields an absolute magnitude of $M_V\simeq0.54$ \citep{vl07,fe08}.  RR Lyrae's metallicity is similar to variables of its class in IC 1613, namely $[Fe/H]\simeq-1.4$ \citep{do01,be02,fe08}.  Consequently, no metallicity correction is required when evaluating the distance to RR Lyrae variables observed in IC 1613.   It follows that the distance to 14 RR Lyrae variables observed in IC 1613 \citep{do01} is $(m-M)_0=24.33\pm0.08$ (assuming $M_V\simeq0.54$, $E_{B-V}=0.05$).  That is in agreement with the classical Cepheid estimate (Table~\ref{dgalaxies}) and negates a sizeable metallicity effect.  Likewise, the distance to RR Lyrae variables observed in the SMC \citep{so02} is $(m-M)_0=18.87\pm0.13$ ($E_{B-V}=0.08$). 

In sum, the results provide the impetus to question a sizeable metallicity effect \citep[see also][]{ud01,pi04,ma09}.

\begin{deluxetable}{lcccccc}
\tabletypesize{\scriptsize}
\tablecaption{Reddenings for Galaxies and Globular Clusters \label{reds}} 
\tablewidth{0pt}
\tablehead{\colhead{Object} &\colhead{[Fe/H]} &\colhead{Cep I} &\colhead{Cep II} &\colhead{\citet{schl98}} &\colhead{\citet{ha96}} &\colhead{2MASS}}
\startdata
LMC & $-0.3$ & $0.14\pm0.04$ & $0.14\pm0.04$ & - & - & - \\
NGC 6441 & $-0.5$ & - & $0.55\pm0.03$ & $0.63$ & $0.45$ & $0.66\pm0.05$  \\
IC 1613 & $-1.0$ & $0.05\pm0.03$ & - & $0.03$ & - &$0.10\pm0.05$ \\
M54 & $-1.6$ & - & $\simeq0.16$ & $0.15$ & $0.15$ & - \\
M15 & $-2.3$ & - & $\simeq0.14$ & $0.11$ & $0.09$ & $0.14\pm0.04$  \\
\enddata
\tablecomments{Metallicites from \citet{mo06}, \citet{ud01}, and \citep{ha96}. \citet{schl98} reddenings computed via the NED extinction calculator. 2MASS field reddenings calculated by Turner following the prescription highlighted in \citet{tu08}.}
\end{deluxetable}

A possible alternative to invoking the effects of metallicity to explain the observed offset in distance between the inner and outer regions is a changing ratio of selective to total extinction, or to consider the spurious photometric effects inherent to sampling the inner regions of galaxies. For example, Cepheids sampling the inner crowded regions of M33 and M106 exhibit larger scatter than classical Cepheids sampling the outer region (Fig.~\ref{fig4}).  For the inner region of M106 the scatter arises in part from a shifting zero-point between shorter and longer-period Cepheids \citep[see also Fig.17,][]{ma06}. That may arise because of a different extinction law than that adopted or photometric contamination.  The distance to the inner regions of M33, M106, and M101 are approximately $10\%$ smaller than values derived for Cepheids in the outer regions.  The trend may be consistent with that expected from crowding and blending \citep{su99,mo00,mo01}.  \citet{ma06} and \citet{sc09} describe their efforts to assess and mitigate such effects for M106 \& M33, and readers are referred to their studies.  The debate surrounding the impact of blending and crowding is as contentious as that of metallicity, highlighting the importance of continuing efforts analogous to the aforementioned studies.  It is tempting to invoke similarities in distance for Type II and classical Cepheids (Table~\ref{dgalaxies}) as a measure of the effects of metallicity and photometric contamination.  Presently, however, an analysis is hindered by small statistics and large uncertainties.  Determining the dependence of metallicity on Cepheid distances using nearby galaxies (e.g., SMC) shall break the degeneracy and facilitate an assessment regarding the effects of crowding and blending on Cepheids in distant galaxies.  More work is needed here.

\section{Parameterization for Interstellar Extinction}
\label{sextinction}
A new photometric colour excess relation based on {\it VI} photometry is provided here to address concerns surrounding the effect of metallicity on Cepheid variables (\S \ref{smetallicity}) and to facilitate the mapping of interstellar reddening for regions of the Milky Way and other galaxies. A classical Cepheid's colour excess can be closely \textit{approximated} (the instability strip exhibits width) by noting that:
\begin{eqnarray}
\nonumber
E_{B-V}=\alpha \log{P} + \beta (m_{\lambda 1} - m_{\lambda 2}) + \phi
\end{eqnarray}
where $\alpha$, $\beta$, and $\phi$ are co-efficients that can be derived by minimizing the $\chi^2$ statistic for a calibrating data set \citep[Table 1 of][]{ma08}, and $m_{\lambda 1}$ and $m_{\lambda 2}$ are photometric magnitudes in different passbands. The optimum solution is:
\begin{equation}
\label{eqn1}
E_{B-V} = -0.28 \log{P} + 0.74 (V-I) - 0.27 \;,
\end{equation}
which reproduces the calibrating set with an average uncertainty of $\pm0.03$ magnitude. The true scatter applying to use of the relationship for individual classical Cepheids shall be larger, particularly for extragalactic variables. {\it VJ}, {\it VH}, and {\it VK} relations are also practical alternatives for determining a colour excess \citep{ma08} and may provide first order estimates to complement reddenings derived by means of {\it BVI$_c$} photometry \citep{ls94,lc07}, spectroscopic analyses \citep{ko08}, and space reddenings \citep{tu84,be07,tu08}.

\section{Summary \& Discussion}
\label{metallicity}
60+ Extragalactic Type II Cepheid candidates were tentatively identified in the galaxies IC 1613, M106, M101, M33, M31, NGC 4603, and the SMC, complementing potential Type II Cepheids found elsewhere in the LMC, NGC 3198, NGC 5128, etc.  A subsequent study examining the extensive DIRECT and CFHT observations for M31 and M33 shall result in a sizeable increase to the statistics \citep{ma04,be08}.  A subsample of potential Type II Cepheids is presented in Table~\ref{arophot}. The list may be contaminated by variables of differing classes which overlap the Wesenheit relation characterizing Type II Cepheids.

The distance established to a set of galaxies from Type II Cepheids agree with literature estimates. That sample includes several galaxies located beyond the Local Group \citep[e.g., M106,][]{ma06}.  Presently, the uncertainties are large and identifications preliminary, yet the results are encouraging and underscore a pertinent role for Type II Cepheids.

Type II Cepheids were likely purged from the final published results of classical Cepheid studies owing to their spurious positions on classical Cepheid period-magnitude relations. A re-examination of the original data sets may indeed be rewarding.  Dedicated surveys of Type II Cepheids (\& RR Lyrae variables) in nearby galaxies are also anticipated, analogous to that conducted of the LMC \citep{ud99,so02,so08}.  As demonstrated in \S \ref{sdistance}, the continued discovery and subsequent analysis of Type II Cepheids, RR Lyrae variables, and classical Cepheids at a common zero-point shall place direct constraints on the effects of metallicity (e.g., SMC and IC 1613).  So too will the discovery and subsequent establishment of mean $VI$ photometry for Type II Cepheids and RR Lyrae variables in globular clusters \citep{cl01,pr03,ho05,mat06,ran07,rab07,co08}. A forthcoming study shall describe how such efforts are being pursued from the Abbey-Ridge Observatory (ARO) \citep{la07,ma08b,tu09c}.   Ongoing photometric monitoring from the ARO, in harmony with archival photometry from the Harvard College Observatory Plate Stacks \citep{gr09}, shall also enable the period evolution of these stars to be ascertained to support evolutionary models \citep{wa02,tu06}.  A holistic approach is needed and modest telescopes may serve a constructive role \citep{pe80,pe86,sz03,pa06,tu09c}.

\subsection*{acknowledgements}
\small{We are grateful to V. Scowcroft, L. Macri, D. Kelson, P. Stetson, X. Bonanos, J. Newman, E. Maoz, J. Hartman, D. Bersier, A. Dolphin, A. Udalski \& I. Soszy{\'n}ski (OGLE),  whose comprehensive surveys were the foundation of the research, to the AAVSO and M. Saladyga, les individus au Centre de Donn\'ees astronomiques de Strasbourg et NASA ADS, L. Berdnikov, L. Szabados, and the RASC. The following reviews and books by \citet{wc84}, \citet{fm96}, \citet{fe69}, \citet{fe99,fe01,fe08b}, \citet{wa02}, and \citet{sz06,sz06b}, were useful in the preparation of this work.} 

\begin{deluxetable}{lllllc}
\tabletypesize{\scriptsize}
\tablecaption{A Sample of Variables Lying Near the Type II Cepheid Wesenheit Relation.\label{arophot}} 
\tablewidth{0pt}
\tablehead{\colhead{Galaxy} &\colhead{Star ID} &\colhead{P (days)} &\colhead{$V$} &\colhead{$I$} &\colhead{Ref.}}
\startdata
IC 1613 & [UWP2001] 10421 & 29.31 &21.099 &20.389 & (1) \\
IC 1613 & [UWP2001] 17473 & 13.12 &21.887 &21.498 & (1) \\
M33 & 118019 & 11.99021 & 22.0394	&21.5377	& (2) \\
M33 & D33J013348.9+304823.0	&77 (SR?) &20.83	&19.16 & (3) \\
M33 & D33J013402.5+302907.5	&89.1 (SR?) &21.52 &19.7 & (3) \\
M33 & D33J013410.6+303750.9	&89.1 &20.77	&19.55 & (3) \\
M33 & D33J013357.1+304455.2	&69.4 &21.47	&20.02 & (3) \\
M33 & D33J013427.3+304407.3	&76.6 &21.46	&20.26 & (3) \\
M31	&	D31J004338.6+414327.8	&	138.4659	(SR?) &	20.16	&	18.5 & (4) \\
M31	&	D31J004443.9+414642.5	&	40.4597	&	21.39	&	19.83 & (4) \\
M31	&	D31J004421.5+413636.1	&	45.1998	&	18.94	&	18.43 & (4) \\
M31	&	D31J004344.8+413705.8	&	31.9068	&	21.47	&	20.1 & (4) \\
M31	&	D31J004454.4+414528.6	&	32.6462	&	21.53	&	20.21 & (4) \\
M31	&	D31J004432.7+414601.0	&	51.7285	&	19.87	&	19.22 & (4) \\ 
M31	&	D31J004336.6+415207.8	&	31.9668	&	19.31	&	19.01 & (4) \\
M31	&	D31J004440.8+413346.0	&	42.6421	&	21.1	&	20.12 & (4) \\ 
M31	&	D31J004306.6+414734.7	&	44.067	&	21.26	&	20.26 & (4) \\ 
M31	&	D31J004426.5+414210.5	&	26.9212	&	21.87	&	20.66 & (4) \\
M31	&	D31J004339.4+413925.8	&	27.4462	&	21.9	&	20.72 & (4) \\
M31	&	D31J004435.4+414237.8	&	33.1692	&	21.18	&	20.33 & (4) \\
M31	&	D31J004441.2+413700.4	&	40.2412	&	20.65	&	20.03 & (4) \\
M31	&	D31J004454.4+415133.9	&	34.374	&	21.22	&	20.44 & (4) \\
M31	&	D31J004325.7+414338.1	&	12.3263	&	21.59	&	20.69 & (4) \\
M101	&	[SSF98] V42	&	54.4	&	24.53	&	23.73	& (5) \\
M101	&	[SSF98] V119	&	71.03	&	23.85	&	23.15	& (5) \\
M101	&	[SSF98] V223	&	95.72	 (SR?) &	24.81	&	23.68	& (5) \\
M101	&	[SSF98] V102	&	72.33	&	25.53	&	24.48	& (5) \\
M101	&	[SSF98] V209	&	94.01 (SR?)	&	24.38	&	23.69	& (5) \\
SMC	&	OGLE SMC-SC2 81443	&	1.2356	 &	19.088	&	18.454	& (6) \\
SMC	&	OGLE SMC-SC5 235485	&	2.11321	&	18.728	&	18.012	&  (6) \\
SMC	&	OGLE SMC-SC3 130452	&	1.48964 (AC?) &	18.734	&	17.993	& (6) \\
SMC	&	OGLE SMC-SC11 100	&	1.88761	&	18.471	&	17.763	& (6) \\
SMC	&	OGLE SMC-SC8 148923	&	1.87772	&	18.127	&	17.511	& (6) \\
SMC	&	OGLE SMC-SC3 157235	&	2.97155	&	18.499	&	17.688	& (6) \\
SMC	&	OGLE SMC-SC5 111664	&	2.56919	&	17.21	&	16.897	& (6) \\
SMC	&	OGLE SMC-SC8 3848	&	3.38938	&	17.29	&	16.776	& (6) \\
SMC	&	OGLE SMC-SC7 83050	&	14.1664	&	16.432	&	15.762	& (6) \\
M106	&	[MSB2006]	I-117359	&	15.69	&	26.987	&	25.927	& (7) \\
M106	&	[MSB2006]	I-139045	&	26.66	&	26.317	&	25.302	& (7) \\
M106	&	[MSB2006]	I-098414	&	20.38	&	26.468	&	25.564	& (7) \\
M106	&	[MSB2006]	I-088850	&	37.64	&	25.911	&	24.862	& (7) \\
M106	&	[MSB2006]	I-071968	&	41.44	&	25.333	&	24.561	& (7) \\
M106	&	[MSB2006]	O-31291 	&	28.62	&	26.602	&	25.538	& (7) \\
M106	&	[MSB2006]	O-38462 	&	14.22	&	26.481	&	25.899	& (7) \\
M106	&	[MSB2006]	I-139786	&	31.75	&	26.349	&	25.372	& (7) \\
M106	&	[MSB2006]	I-078417	&	33.31	&	26.395	&	25.417	& (7) \\
M106	&	[MSB2006]	O-07822 	&	39.55	&	25.855	&	25.004	& (7) \\
M106	&	[MSB2006]	I-005860	&	36.7	&	25.857	&	25.067	& (7) \\
M106	&	[MSB2006]	I-239712	&	24.31	&	26.849	&	25.906	& (7) \\
M106	&	[MSB2006]	I-052900	&	42.4	&	25.634	&	24.925	& (7) \\
M106	&	[MSB2006]	I-120571	&	42.01	&	25.913	&	25.09 & (7) \\
M106	&	[MSB2006]	O-28609 	&	35.69	&	26.439	&	25.51 & (7) \\
M106	&	[MSB2006]	I-095468	&	29.86	&	26.706	&	25.781	& (7) \\ 
M106	&	[MSB2006]	O-29582 	&	30.92	&	26.605	&	25.706	& (7) \\
M106	&	[MSB2006]	I-106574	&	29.2	&	26.442	&	25.702	& (7) \\
M106	&	[MSB2006]	I-139636	&	41.12	&	26.411	&	25.456	& (7) \\
M106	&	[MSB2006]	O-11134 	&	38.33	&	26.075	&	25.333	& (7) \\
M106	&	[MSB2006]	I-082122	&	43.16	&	26.305	&	25.387	& (7) \\
NGC 4603	&	1165	&	60	&	27.24	&	26.53	&  (8) \\
NGC 4603	&	2848	&	59	&	26.9	&	26.56	&  (8) \\
NGC 4603	&	2862	&	33	&	27.72	&	27.28	& (8) \\
NGC 4603	&	2547	&	26	&	27.57	&	27.23	& (8) \\
NGC 4603	&	1545	&	25	&	27.55	&	27.22	& (8) \\
\enddata
\tablecomments{References: (1) \citet{ud01}, (2) \citet{sc09}, (3) \citet{ma01}, (4) \citet{bo03}, (5) \citet{st98}, (6) \citet{ud99}, (7) \citet{ma06}, (8) \citet{new01}. \\}
\end{deluxetable}


\begin{thebibliography}{}
\bibitem[Andrievsky et al.(2002)]{an02} Andrievsky S.~M. et al., 2002, A\&A, 381, 32 
\bibitem[Andrievsky et al.(2002b)]{an02b} Andrievsky, S.~M., Kovtyukh, V.~V., Luck, R.~E., L{\'e}pine, J.~R.~D., Maciel, W.~J., \& Beletsky, Y.~V.\ 2002, A\&A, 392, 491 
\bibitem[Benedict et al.(2002)]{be02} Benedict G.~F. et al., 2002, AJ, 123, 473 
\bibitem[Benedict et al.(2007)]{be07} Benedict G.~F. et al., 2007, AJ, 133, 1810 
\bibitem[Bersier et al.(2008)]{be08} Bersier, D., et al.\ 2008, arXiv:0803.2010 
\bibitem[Berdnikov \& Turner(2001)]{be01} Berdnikov, L.~N., \& Turner, D.~G.\ 2001, ApJS, 137, 209 
\bibitem[Bird et al.(2009)]{bi09} Bird, J.~C., Stanek, K.~Z., \& Prieto, J.~L.\ 2009, \apj, 695, 874 
\bibitem[Bonanos et al.(2003)]{bo03} Bonanos, A.~Z., Stanek, K.~Z., Sasselov, D.~D., Mochejska, B.~J., Macri, L.~M., \& Kaluzny, J.\ 2003, AJ, 126, 175 
\bibitem[Caldwell \& Coulson(1985)]{cc85} Caldwell, J.~A.~R., \& Coulson, I.~M.\ 1985, \mnras, 212, 879 
\bibitem[Chiosi et al.(1993)]{ch93} Chiosi, C., Wood, P.~R., \& Capitanio, N.\ 1993, ApJS, 86, 541 
\bibitem[Clementini et al.(2003)]{cl03} Clementini, G., Gratton, R., Bragaglia, A., Carretta, E., Di Fabrizio, L., \& Maio, M.\ 2003, AJ, 125, 1309 
\bibitem[Clement et al.(2001)]{cl01} Clement, C.~M., et al.\ 2001, AJ, 122, 2587 
\bibitem[Corwin et al.(2008)]{co08} Corwin, T.~M., Borissova, J., Stetson, P.~B., Catelan, M., Smith, H.~A., Kurtev, R., \& Stephens, A.~W.\ 2008, AJ 135, 1459 
\bibitem[Cutri et al.(2003)]{cu03} Cutri R.~M. et al., 2003, The IRSA 2MASS All-Sky Point Source Catalog of Point Sources, NASA/IPAC Infrared Science Archive 
\bibitem[Dolphin et al.(2001)]{do01} Dolphin, A.~E., et al.\ 2001, \apj, 550, 554 
\bibitem[Feast(1999)]{fe99} Feast M., 1999, PASP, 111, 775 
\bibitem[Feast(2001)]{fe01} Feast M., 2001, arXiv:astro-ph/0110360 
\bibitem[Feast et al.(2008)]{fe08} Feast M.~W., Laney C.~D., Kinman T.~D., van Leeuwen F., Whitelock P.~A., 2008, MNRAS, 386, 2115
\bibitem[Feast(2008)]{fe08b} Feast, M.~W.\ 2008, arXiv:0806.3019 
\bibitem[Fernie(1969)]{fe69} Fernie, J.~D.\ 1969, \pasp, 81, 707 
\bibitem[Fernie(1976)]{fe76} Fernie D.\ 1976, The Whisper and the Vision - The Voyages of the Astronomers.  
\bibitem[Fernie(2002)]{fe02} Fernie J.~D., 2002, Setting sail for the universe : astronomers and their discoveries, Rutgers University Press, New Brunswick, NJ
\bibitem[Ferrarese et al.(2007)]{fe07} Ferrarese L., Mould J.~R., Stetson P.~B., Tonry J.~L., Blakeslee J.~P., Ajhar E.~A., 2007, ApJ, 654, 186 
\bibitem[Fouqu{\'e} et al.(2007)]{fo07} Fouqu{\'e} P. et al., 2007, A\&A, 476, 73
\bibitem[Freedman et al.(2001)]{fr01} Freedman W.~L. et al., 2001, ApJ, 553, 47
\bibitem[Freedman \& Madore(1996)]{fm96} Freedman, W.~L., \& Madore, B.~F.\ 1996, Clusters, Lensing, and the Future of the Universe, 88, 9 
\bibitem[Gascoigne(1974)]{ga74} Gascoigne, S.~C.~B.\ 1974, \mnras, 166, 25P 
\bibitem[Gieren et al.(1999)]{gi99} Gieren, W.~P., Moffett, T.~J., \& Barnes, T.~G., III 1999, \apj, 512, 553 
\bibitem[Gieren et al.(2008)]{gi08} Gieren, W., Pietrzy{\'n}ski, G., Soszy{\'n}ski, I., Bresolin, F., Kudritzki, R.-P., Storm, J., \& Minniti, D.\ 2008, ApJ, 672, 266 
\bibitem[Grindlay et al.(2009)]{gr09} Grindlay, J., Tang, S., Simcoe, R., Laycock, S., Los, E., Mink, D., Doane, A., \& Champine, G.\ 2009, Astronomical Society of the Pacific Conference Series, 410, 101 
\bibitem[Groenewegen et al.(2008)]{gr08} Groenewegen M.~A.~T., Udalski A., Bono, G., 2008, A\&A, 481, 441
\bibitem[Harris \& Welch(1989)]{hw89} Harris, H.~C., \& Welch, D.~L.\ 1989, \aj, 98, 981
\bibitem[Harris(1996)]{ha96} Harris, W.~E.\ 1996, AJ, 112, 1487 
\bibitem[Hartman et al.(2006)]{ha06} Hartman, J.~D., Bersier, D., Stanek, K.~Z., Beaulieu, J.-P., Kaluzny, J., Marquette, J.-B., Stetson, P.~B., \& Schwarzenberg-Czerny, A.\ 2006, MNRAS, 371, 1405 
\bibitem[Herrnstein et al.(2005)]{he05} Herrnstein J.~R., Moran J.~M., Greenhill L.~J., Trotter A.S., 2005, ApJ, 629, 719
\bibitem[Hoffleit(2002)]{ho02} Hoffleit, D.\ 2002, Misfortunes as blessings in disguise : the story of my life, by Dorrit Hoffleit Cambridge, MA: American Association of Variable Star Observers (AAVSO), 2002.,  
\bibitem[Horne(2005)]{ho05} Horne, J.~D.\ 2005, Journal of the American Association of Variable Star Observers (JAAVSO), 34, 61 
\bibitem[Kelson et al.(1996)]{ke96} Kelson, D.~D., et al., 1996, ApJ, 463, 26 
\bibitem[Kelson et al.(1999)]{ke99} Kelson D.~D., et al., 1999, ApJ, 514, 614 
\bibitem[Kovtyukh et al.(2008)]{ko08} Kovtyukh V.~V., Soubiran C., Luck R.~E., Turner D.~G., Belik S.~I., Andrievsky S.~M., Chekhonadskikh F.~A., 2008, MNRAS, 389, 1336 
\bibitem[Kubiak \& Udalski(2003)]{ku03} Kubiak M., Udalski A., 2003, Acta Astr., 53, 117 
\bibitem[Laney \& Stobie(1994)]{ls94} Laney C.~D., Stobie R.~S., 1994, MNRAS, 266, 441 
\bibitem[Laney \& Caldwell(2007)]{lc07} Laney C.~D., Caldwell J.~A.~R., 2007, MNRAS, 377, 147 
\bibitem[Lane(2007)]{la07} Lane D.~J., 2007, 96th Spring Meeting of the AAVSO, http://www.aavso.org/aavso/meetings/spring07present/Lane.ppt
\bibitem[Luck et al.(1998)]{lu98} Luck R.~E., Moffett T.~J., Barnes T.~G., Gieren W.~P., 1998, AJ, 115, 605 
\bibitem[Madore(1982)]{ma82} Madore B.~F., 1982, ApJ, 253, 575
\bibitem[Madore \& Freedman(1991)]{mf91} Madore B.~F., Freedman W.~L., 1991, PASP, 103, 933
\bibitem[Macri et al.(2001)]{ma01} Macri, L.~M., Stanek, K.~Z., Sasselov, D.~D., Krockenberger, M., \& Kaluzny, J.\ 2001, AJ, 121, 870 
\bibitem[Macri et al.(2001b)]{ma01b} Macri, L.~M., et al.\ 2001 (b), ApJ, 549, 721 
\bibitem[Macri(2004)]{ma04} Macri, L.~M.\ 2004, IAU Colloq.~193: Variable Stars in the Local Group, 310, 33 
\bibitem[Macri et al.(2006)]{ma06} Macri, L.~M., Stanek, K.~Z., Bersier, D., Greenhill, L.~J., \& Reid, M.~J.\ 2006, ApJ, 652, 1133 
\bibitem[Majaess et al.(2008)]{ma08} Majaess D.~J., Turner D.~G., Lane D.~J., 2008, MNRAS, 390, 1539
\bibitem[Majaess et al.(2008b)]{ma08b} Majaess D.~J., Turner D.~G., Lane D.~J., Moncrieff K.~E., 2008 (b), JAAVSO, 36, 90 
\bibitem[Majaess et al.(2009)]{ma09} Majaess, D.~J., Turner, D.~G., \& Lane, D.~J.\ 2009, MNRAS, 398, 263 
\bibitem[Matsunaga et al.(2006)]{mat06} Matsunaga, N., et al.\ 2006, MNRAS, 370, 1979 
\bibitem[Matsunaga et al.(2009)]{mat09} Matsunaga, N., Feast, M.~W., \& Menzies, J.~W.\ 2009, MNRAS, 397, 933 
\bibitem[Maoz et al.(1999)]{mao99} Maoz, E., Newman, J.~A., Ferrarese, L., Stetson, P.~B., Zepf, S.~E., Davis, M., Freedman, W.~L., \& Madore, B.~F.\ 1999, Nature, 401, 351 
\bibitem[Mochejska et al.(2000)]{mo00} Mochejska, B.~J., Macri, L.~M., Sasselov, D.~D., \& Stanek, K.~Z.\ 2000, AJ, 120, 810 
\bibitem[Mochejska et al.(2001)]{mo01} Mochejska, B.~J., Macri, L.~M., Sasselov, D.~D., \& Stanek, K.~Z.\ 2001, arXiv:astro-ph/0103440 
\bibitem[Mottini(2006)]{mo06} Mottini M., 2006, Ph.D.~Thesis
\bibitem[Newman et al.(1999)]{ne99} Newman, J.~A., Zepf, S.~E., Davis, M., Freedman, W.~L., Madore, B.~F., Stetson, P.~B., 
Silbermann, N., \& Phelps, R.\ 1999, ApJ, 523, 506 
\bibitem[Newman et al.(2001)]{new01} Newman, J.~A., Ferrarese, L., Stetson, P.~B., Maoz, E., Zepf, S.~E., Davis, M., Freedman, 
W.~L., \& Madore, B.~F.\ 2001, ApJ, 553, 562 
\bibitem[Opolski(1983)]{op83} Opolski A., 1983, IAU Inf. Bull. Var. Stars, 2425, 1 
\bibitem[Paczy{\'n}ski(2006)]{pa06} Paczy{\'n}ski, B.\ 2006, \pasp, 118, 1621 
\bibitem[Percy(1980)]{pe80} Percy J.~R., 1980, JRASC, 74, 334
\bibitem[Percy(1986)]{pe86} Percy, J.~R.\ 1986, Study of Variable Stars using Small Telescopes,  
\bibitem[Pellerin et al.(2009)]{pe09} Pellerin, A., Macri, L.~M., Bradshaw, A.~K., 
\& Stanek, K.~Z.\ 2009, American Institute of Physics Conference Series, 1170, 40 
\bibitem[Pietrzy{\'n}ski et al.(2004)]{pi04} Pietrzy{\'n}ski, G., Gieren, W., Udalski, A., Bresolin, F., Kudritzki, R.-P., Soszy{\'n}ski, I., Szyma{\'n}ski, M., 
\& Kubiak, M.\ 2004, AJ, 128, 2815 
\bibitem[Pietrzy{\'n}ski et al.(2006)]{pi06} Pietrzy{\'n}ski, G., et al.\ 2006, AJ, 132, 2556 
\bibitem[Pritzl et al.(2003)]{pr03} Pritzl B.~J., Smith H.~A., Stetson P.~B., Catelan M., Sweigart A.~V., Layden A.~C., Rich R.~M., 2003, AJ, 126, 1381 
\bibitem[Rabidoux et al.(2007)]{rab07} Rabidoux, K., et al.\ 2007, Bulletin of the American Astronomical Society, 38, 845 
\bibitem[Randall et al.(2007)]{ran07} Randall, J.~M., Rabidoux, K., Smith, H.~A., De Lee, N., Pritzl, B., \& Osborn, W.\ 2007, Bulletin of the American Astronomical Society, 38, 276 
\bibitem[Schlegel et al.(1998)]{schl98} Schlegel, D.~J., Finkbeiner, D.~P., \& Davis, M.\ 1998, ApJ, 500, 525 
\bibitem[Schmidt et al.(2004)]{es04} Schmidt, E.~G., Johnston, D., Langan, S., \& Lee, K.~M.\ 2004, AJ, 128, 1748 
\bibitem[Schmidt et al.(2005c)]{es05c} Schmidt, E.~G., Johnston, D., Langan, S., \& Lee, K.~M.\ 2005 (c), AJ, 130, 832 
\bibitem[Schmidt et al.(2005b)]{es05b} Schmidt, E.~G., Johnston, D., Langan, S., \& Lee, K.~M.\ 2005 (b), AJ, 129, 2007 
\bibitem[Schmidt et al.(2005)]{es05} Schmidt, E.~G., Langan, S., Rogalla, D., \& Thacker-Lynn, L.\ 2005, Bulletin of the American Astronomical Society, 37, 1334 
\bibitem[Schmidt et al.(2009)]{es09} Schmidt, E.~G., Hemen, B., Rogalla, D., \& Thacker-Lynn, L.\ 2009, AJ, 137, 4598 
\bibitem[Scowcroft et al.(2009)]{sc09}  Scowcroft, V., Bersier, D., Mould, J.~R., \& Wood, P.~R.\ 2009, MNRAS, 396, 1287 
\bibitem[Sebo et al.(2002)]{se02} Sebo, K.~M., et al.\ 2002, \apjs, 142, 71 
\bibitem[Soszynski et al.(2002)]{so02} Soszynski, I., et al.\ 2002, Acta Astronomica, 52, 369 
\bibitem[Soszynski et al.(2007)]{so07} Soszynski, I., et al.\ 2007, Acta Astronomica, 57, 201 
\bibitem[Soszynski et al.(2008)]{so08} Soszy{\'n}ski, I., et al.\ 2008, Acta Astronomica, 58, 293 
\bibitem[Soszynski et al.(2009)]{so09} Soszy{\~n}ski, I., et al.\ 2009, Acta Astronomica, 59, 239 
\bibitem[Stanek et al.(1998)]{sta98} Stanek, K.~Z., Zaritsky, D., \& Harris, J.\ 1998, ApJL, 500, L141 
\bibitem[Stanek \& Udalski(1999)]{su99} Stanek, K.~Z., \& Udalski, A.\ 1999, arXiv:astro-ph/9909346 
\bibitem[Stetson et al.(1998)]{st98} Stetson, P.~B., et al.\ 1998, ApJ, 508, 491 
\bibitem[Szabados(2003)]{sz03} Szabados, L.\ 2003, Astrophysics and Space Science Library, 289, 207 
\bibitem[Szabados(2006)]{sz06} Szabados, L.\ 2006, Commmunications of the Konkoly Observatory Hungary, 104, 105 
\bibitem[Szabados(2006b)]{sz06b} Szabados, L.\ 2006 (b), Odessa Astronomical Publications, 18, 111 
\bibitem[Tammann et al.(2003)]{ta03} Tammann, G.~A., Sandage, A., \& Reindl, B.\ 2003, \aap, 404, 423 
\bibitem[Thim et al.(2003)]{th03} Thim, F., Tammann, G.~A., Saha, A., Dolphin, A., Sandage, A., Tolstoy, E., \& Labhardt, L.\ 2003, ApJ, 590, 256 
\bibitem[Turner(1984)]{tu84} Turner, D.~G.\ 1984, JRASC, 78, 229 
\bibitem[Turner(1996)]{tu96b} Turner D.~G., 1996, JRASC, 90, 82
\bibitem[Turner(2001)]{tu01} Turner D.~G., 2001, Odessa Astr. Publ., 14, 166 
\bibitem[Turner \& Burke(2002)]{tu02} Turner D.~G., Burke J.~F., 2002, AJ, 124, 2931
\bibitem[Turner et al.(2005)]{tu05} Turner D.~G., Savoy J., Derrah J., Abdel-Sabour Abdel-Latif M., Berdnikov L.~N., 2005, PASP, 117, 207
\bibitem[Turner et al.(2006)]{tu06} Turner D.~G., Abdel-Sabour Abdel-Latif M., Berdnikov L.~N., 2006, PASP, 118, 410
\bibitem[Turner et al.(2008)]{tu08} Turner D.~G., MacLellan R. F., Henden A. A., Berdnikov L. N., 2008, PASP, submitted
\bibitem[Turner(2009)]{tu09b} Turner D.~G., 2009, submitted.
\bibitem[Turner et al.(2009)]{tu09c} Turner, D.~G., Majaess, D.~J., Lane, D.~J., Szabados, L., Kovtyukh, V.~V., Usenko, I.~A., \& Berdnikov, L.~N.\ 2009, arXiv:0907.2969 
\bibitem[Udalski et al.(1998)]{ud98} Udalski, A., Szymanski, M., Kubiak, M., Pietrzynski, G., Wozniak, P., \& Zebrun, K.\ 1998, Acta Astronomica, 48, 1 
\bibitem[Udalski et al.(1999)]{ud99} Udalski A. et al., 1999, Acta Astr., 49, 223
\bibitem[Udalski et al.(2001)]{ud01} Udalski, A., Wyrzykowski, L., Pietrzynski, G., Szewczyk, O., Szymanski, M., Kubiak, M., Soszynski, I., \& Zebrun, K.\ 2001, Acta Astronomica, 51, 221 
\bibitem[Udalski(2003)]{ud03} Udalski, A.\ 2003, ApJ, 590, 284 
\bibitem[van den Bergh(1968)]{vb68} van den Bergh S., 1968, JRASC, 62, 145
\bibitem[van Leeuwen et al.(2007)]{vl07} van Leeuwen, F., Feast, M.~W., Whitelock, P.~A., \& Laney, C.~D.\ 2007, MNRAS, 379, 723 
\bibitem[Wallerstein \& Cox(1984)]{wc84} Wallerstein, G., \& Cox, A.~N.\ 1984, \pasp, 96, 677 
\bibitem[Wallerstein(2002)]{wa02} Wallerstein, G.\ 2002, PASP, 114, 689 
\bibitem[Zepf et al.(1997)]{ze97} Zepf, S.~E., Newman, J., Davis, M., Freedman, W.~L., Madore, B.~F., \& Silbermann, N.~A.\ 1997, Bulletin of the American Astronomical Society, 29, 1208 
\end{thebibliography}
\end{document}